\documentstyle[preprint,aps,epsfig]{revtex}
\def\be{\begin{equation}}
\def\ee{\end{equation}}
\begin{document}
\draft\title{Heating the intergalactic medium by radiative decay of 
neutrinos} 
\author{M. H. Chan and M.-C. Chu}
\address{Department of Physics, The Chinese University of Hong Kong, 
\\  Shatin, New Territories, Hong Kong, China}
\maketitle
\begin{abstract}
Assuming that there exists a species of heavy sterile neutrinos 
($m_{\nu}>1$ keV) and their decays can serve as a heating source of the 
hot gas in galaxy clusters, we study how the observational constraints on 
cooling flow limit the mass of these sterile neutrinos. We
predict a relation among the luminosity, total mass and the 
redshift of a cluster, and we compare this relation with data from 12 
clusters to obtain an estimate of the decay rate of the sterile neutrino.
\end{abstract} 
\vskip 5mm

\section{Introduction}
Recent observations indicate that there are not enough cooling flows in 
galaxy clusters to explain why hot gas in some clusters  
are so hot \cite{coolingflow}, \cite{coolingflow2}. On the other hand, 
neutrino oscillation experiments (solar, atmospheric and LSND experiments) 
indicate that 3 ordinary neutrinos cannot explain 
the three different scales of the mass squared differences between 
neutrino species $\Delta m^2$'s, of about $10^{-5}$ eV$^2$, 
$10^{-2}$ eV$^2$ and 1 eV$^2$ \cite{oscillation}, suggesting that one or 
more sterile neutrinos must be added. In this article, we study the 
possibility that a species of massive sterile neutrinos ($m_{\nu}>1$ keV) 
exists and their 
decays provide a heating source to maintain the temperature 
of the hot gas. We use this model to predict a relation among 
the luminosity, total mass and redshift of a cluster, and we obtain an 
estimate of the decayrate 
of the sterile neutrinos by using available data from 12 clusters. This 
idea of the radiative decay of neutrinos as an 
energy source has been proposed to account for reionization \cite{reion}. 

In standard cosmology, neutrinos with $m_{\nu}<1$ MeV are classified 
as hot dark matter (HDM) because they are relativistic at decoupling. As 
the universe expands, they cool down and become non-relativistic after a 
redshift of 
\begin{equation}
z_{rel} \approx \frac{m_{\nu}c^2}{3kT_{\nu,0}}=2 \times 10^3 \left( 
\frac{m_{\nu}}{1~\rm{eV}} \right),
\end{equation}
where $T_{\nu,0}$ is the neutrino temperature today. 
In the non-relativistic regime, the 
average neutrino speed at neutrino temperature $T_{\nu}$ is \cite{Chung}
\begin{equation}
<v>=160~{\rm{km/s}} \left( \frac{1~{\rm{eV}}}{m_{\nu}} \right) \left( 
\frac{T_{\nu}}{1.947~\rm{K}} \right).
\end{equation}
They do not contribute to structure formation if their average speed is 
too high. For example, at matter-radiation equality (redshift $\approx 
10^4 h^2$, $h$ being the Hubble parameter), the 
neutrino speed is close to $c$ if $m_{\nu} \ll 10$ eV. 
By using perturbation theory on neutrino clustering, it was found 
that massive 
neutrinos can gravitationally cluster if $m_{\nu}$ is greater than 
several eV's \cite{Chung}. Therefore, neutrinos with mass $m_{\nu} \ge$10 
eV can be classified as warm dark matter (WDM) rather than HDM.

\section{Sterile neutrino decay in clusters}
In this model, we assume that all 3 types of active neutrinos with masses 
$m_1, m_2$ and $m_3$ did not contribute to structure formation and are 
thus classified as HDM \cite{m1}:
\begin{equation}
\frac{m_1+m_2+m_3}{94~\rm{eV}}= \Omega_{\nu}h^2,
\end{equation}
where $\Omega_{\nu}$ is the light neutrino density parameter.
In addition, we assume that all cold dark matter are composed of two types 
of sterile neutrinos 
with masses $m_{s1}$ and $m_{s2}$, and the latter will decay 
into $m_1$ neutrinos with decay rate $\Gamma$:
\begin{equation}
\nu_{s2} \rightarrow \nu_1+ \gamma.
\end{equation}

The massive sterile neutrinos may be produced 
in the early universe by oscillations from active neutrinos 
\cite{Dodelson}. From the standard cosmology, the number densities of 
sterile neutrinos $n_{s1}$ and $n_{s2}$ at present (time = $t_0$) are 
upper bounded by \cite{reion}
\begin{equation}
n_{s1}=1.2\times 10^{-2} n_0 \left(\frac{\rm{1~keV}}{m_{s1}}\right) 
\left(\frac{\Omega_{s1}}{0.23}\right) \left(\frac{h^2}{0.72}\right)= 
\frac{\alpha_1}{m_{s1}}n_0, 
\end{equation}
\begin{equation}
n_{s2}=1.2\times 10^{-2} n_0 \left(\frac{\rm{1~keV}}{m_{s2}}\right) 
\left(\frac{\Omega_{s2}}{0.23}\right) \left(\frac{h^2}{0.72}\right)= 
\frac{\alpha_2}{m_{s2}}n_0e^{-\Gamma t_0}, 
\end{equation}
where $\alpha_1<m_{s1}$ and $\alpha_2<m_{s2}$ are constants which are 
dependent on the cosmological sterile neutrino 
density parameters $\Omega_{s1}$ and $\Omega_{s2}$ respectively, and $n_0$ 
is the number density of one type of the active neutrinos now ($n_0 
\approx 100$ cm$^{-3}$ if there were no sterile neutrino 
decays \cite{number}). If the 
sterile neutrinos can decay to $m_1$ neutrinos emitting a photon 
with energy $\epsilon=(m_{s2}-m_1)c^2 \approx m_{s2}c^2$, the number 
density of the $m_1$ neutrinos now is given by
\begin{equation}
n_1=n_0 \left[1+ \frac{\alpha_2}{m_{s2}}(1-e^{-\Gamma t_0}) \right].
\end{equation}
Therefore, we can write the cold dark matter density parameter $\Omega_m$ 
as
\begin{equation}
\Omega_m=\frac{\alpha_1+\alpha_2e^{-\Gamma t_0}}{94~\rm{eV}}.
\end{equation}

The interaction of the radiated photons and 
electrons in the cluster hot gas can be treated as a 
heating 
source of the hot gas. There exists a lower limit of the energy of the 
photons such that they 
can undergo Compton Scattering with electrons rather than Inverse Compton 
Scattering (see Appendix). The energy of photons must be greater than the 
energy of the electrons in order to transfer energy to the hot gas, which 
typically has temperature of about several keV's in  
clusters. Therefore, in our model, $m_{s2}$ must be at least of keV 
order. Here, we set $m_1=m_2=m_3$, as  
neutrino oscillation experiments indicate that the difference in the 
mass squared of three active neutrino species,
$\Delta m^2$'s, are about $10^{-5}$ eV and $10^{-2}$ eV \cite{mass}.  

Eq.~(8) gives one relation between two unknowns, $\alpha_1$ and 
$\alpha_2$. 
Another relation comes from the decay of $m_{s2}$ neutrinos, which we 
assume to be 
the source of hot gas luminosity. The total non-baryonic mass in a cluster 
is given by
\begin{equation}
M=M_1+M_2+M_3+M_{s1}+M_{s2},
\end{equation}
where $M_1$, $M_2$, $M_3$, $M_{s1}$ and $M_{s2}$ are the total masses of 
three types of active neutrinos and two types of sterile neutrinos 
respectively. By combining Eqs.~(5), (6), (7) and (9), we get
\begin{equation}
N_{s2}=\frac{M}{m_1} \left[\frac{\alpha_2e^{-\Gamma t_0}}{3r_{s2}m_1+ 
\alpha_2r_{s2}e^{-\Gamma t_0}+ \alpha_2(1-e^{-\Gamma t_0})+ 
\alpha_1r_{s2}} \right],
\end{equation}
where $N_{s2}$ is the total number of sterile neutrinos inside a cluster, 
and $r_{s2}=m_{s2}/m_1$. The total power given out by the radiative 
neutrinos in a cluster at time $t$ 
is $L_x=N_{s2} \Gamma \epsilon=N_{s2} \Gamma r_{s2}m_1c^2$. Therefore we 
get
\begin{equation}
L_x=M \Gamma c^2 \left[\frac{ \alpha_2 e^{-\Gamma t}}{3m_1+ 
\alpha_2 e^{-\Gamma t}+ \alpha_2(1-e^{-\Gamma t})/r_{s2}+ \alpha_1} 
\right].
\end{equation}

At the time of CMB last scattering ($t_l \approx 10^{13}$ s), the cold 
dark matter density $\rho_l$ is given by
\begin{equation}
\rho_la_l^3=\alpha_1n_0+ \alpha_2n_0e^{- \Gamma t_l},
\end{equation}
where $a_l$ is the scale factor of the universe at $t_l$. 
In order to heat up the hot gas, the value of $\Gamma$ must be greater 
than $3 \times 10^{-17}$ s$^{-1}$ by using Eq.~(11) (also see Fig.~1). 
Therefore the factor $e^{- \Gamma 
t_l} \approx 1$ and $\rho_la_l^3 \approx \alpha_1n_0+ \alpha_2n_0$. At 
the time of galaxy formation ($t_g>10^{17}$ s), the dark matter density 
$\rho_g$ is
\begin{equation}
\rho_ga_g^3=\alpha_1n_0+ \alpha_2n_0e^{- \Gamma t_g},
\end{equation}
where $a_g$ is the scale factor of the universe at $t_g$.
The factor $e^{- \Gamma t_g}<0.03$ is much less than 1, and we 
get $\rho_{g}a_g^3 \approx \alpha_1n_0$. Therefore, at $t_g$, $m_{s1}$ 
neutrinos dominate all the dark matter. We can therefore obtain 
another relation between $\alpha_1$ and $\alpha_2$ by taking the 
difference of Eq.~(12) 
and Eq.~(13). So the density of $m_{s2}$ neutrinos $\rho_{s2}$ at $t_l$ is 
given by
\begin{equation}
\frac{\rho_{s2}a_l^3}{\rho_c}=
\frac{\rho_{l}a^3_l}{\rho_c}- 
\frac{\rho_{g}a^3_g}{\rho_c}= \Omega^{\rm{CMB}}_m- 
\Omega^{\rm{2dFGRS}}_m \approx \frac{ \alpha_2n_0}{\rho_c},
\end{equation} 
where $\rho_c$ is the critical density of the 
universe, $\Omega^{\rm{CMB}}_mh^2=0.14 \pm 0.02$ is 
from WMAP data and $\Omega^{\rm{2dFGRS}}_mh^2=0.134 \pm 0.006$ is 2dFGRS 
data corresponding to CMB last scattering and galaxy formation 
respectively \cite{Data}. The corresponding value of $\alpha_1= 
94 \times \Omega^{\rm{2dFGRS}}_mh^2$ eV = $12.60 \pm 0.56$ eV, which means
that $m_{s1}$ must be greater than 12 eV and consistent 
with our assumption that they can form structures. The upper bound of 
$\alpha_2$ obtained by Eq.~(14) is 
about 3 eV. 

By using these values, we can find the value of $\Gamma$ as well 
as the corresponding mass components in cluster.
Since $\alpha_1$ is much greater than $m_1$, $\alpha_2e^{- \Gamma 
t_0}$ and $\alpha_2/r_{s2}$, we can rewrite Eq.~(11) as 
\begin{equation}
\ln{ \left( \frac{L_x}{M}\right)} \approx \ln{ \left( \frac{\Gamma c^2 
\alpha_2}{\alpha_1} \right)}- \Gamma \left[\frac{t_0}{(1+z)^{3/2}} 
\right],
\end{equation}
where $z$ is the redshift of a cluster and we have assumed matter 
dominated expansion and $r_{s2}$ is much greater than 1.
By using data of redshifts, luminosities and masses from 12 clusters (see 
Table 1), we can estimate $\Gamma$ from the 
slope of $\ln L_x/M$ vs.~$(1+z)^{-3/2}$ (see 
Fig.~1). The slope and y-intercept obtained are $-18.23 \pm 4.19$ and $ 
18.50 \pm 3.87$, which 
imply $\Gamma \sim (4.6 \pm 1.1) \times 10^{-17}$ s$^{-1}$. 

The corresponding mass components in clusters with $i=1,2,3,s1$ or $s2$ 
are given by
\begin{equation}
M_i=\frac{\Omega_i}{\Omega_m}M,
\end{equation}
where $\Omega_i$ is the density parameter of neutrinos with $m_i$.
For a typical cluster with $M \sim 10^{15}M_{\odot}$, the present value of 
$M_{s2}$ is about $10^{6}M_{\odot}$.

\section{Temperature profile of hot gas in a cluster}
We can predict the temperature profile of the hot gas in a cluster
by using a simple energy flow equation. We first look at the mechanism of 
how the energy from neutrinos decays is transferred to the hot gas.

Since the $m_{s2}$ neutrinos have mass greater than keV, they must have 
gravitationally 
collapsed to galactic scale structures \cite{Viollier}, \cite{manho}. The 
radius of 
a hydrostatic `neutrino star' is given by \cite{star}
\begin{equation}
R=20.7~{\rm{pc}} \left( \frac{M_{s2}}{10^6 M_{\odot}} \right)^{-1/3} 
\left( \frac{m_{s2}}{1~\rm{keV}} \right)^{-8/3}.
\end{equation}
From our estimate, $M_{s2}$ and $m_{s2}$ are of order $10^6$ 
solar mass and keV respectively. Therefore, 
the size of a corresponding `sterile neutrino star' is about 20 pc which 
may hide deeply inside the galactic bulge. 

Suppose a photon travels from $r=0$ and 
collides with electrons in a cluster. The number of collisions of a
photon and electrons within a 
radius $r$ is approximately given by the optical depth:
\begin{equation}
x= \int^r_0 \frac{dr'}{\bar{l}}= \int^r_0n_e \sigma dr',
\end{equation}
where $\bar{l}$ is the mean free path of a photon, and $n_e$ and $\sigma$ 
are the electron 
number density and Compton cross section respectively. In a cluster, the 
electron number density is given by
\begin{equation}
n_e= \frac{n_c}{[1+(r/r_c)^2]^{3 \beta_c/2}},
\end{equation}
where $n_c$, $r_c$ and $\beta_c$ are parameters in the cluster beta model 
\cite{data}.
In Fig.~2, we plot the number of collisions $x$ vs.~$r$. We can see that 
$x$ is much less than 1 for all $r$. 
Therefore, a photon does not collide frequently with electrons in 
a cluster. However, inside the Milkyway galactic bulge, 
the number density of electrons is
\begin{equation}
n_e= \frac{n'_c}{[1+(r/r'_c)]^{1.8}},
\end{equation}
where $n'_c$ and $r'_c$ are parameters which are equal to $1.6 \times 
10^{8}$ 
cm$^{-3}$ and 0.34 pc respectively \cite{nature}. Fig.~3 shows the 
number of collisions
inside the galactic bulge. We can see that $x$ is much greater than 1 
for 
small $r$. Therefore, we can believe that the energy of the 
photons are first absorbed 
by the electrons inside the galactic bulge and then the energy is 
transferred to the hot gas by conduction inside the cluster. 
The mean free path for conduction is \cite{xray}
\begin{equation}
\bar{l_c}=23~{\rm{kpc}} \left( \frac{T_g}{10^8~{\rm{K}}} \right)^2 \left( 
\frac{n_e}{10^{-3}~{\rm{cm}}^{-3}} \right)^{-1},
\end{equation}
which is smaller than the length scale of a cluster ($\approx$ 1 Mpc) with 
the temperature of hot gas $T_g \approx 10^8$ K.
The energy flux per unit time for conduction is given by
\begin{equation}
j(r)=-K_c \frac{dT_g}{dr},
\end{equation}
where 
\begin{equation}
K_c=1.31n_e \bar{l_c}k \left( \frac{kT_g}{m_e} \right)^{\frac{1}{2}}.
\end{equation}
We can obtain the temperature profile $T_g(r)$ by solving Eq.~(22) with 
$j(r)=L(r)/4 \pi r^2$ and $L(r)= \int 4 \pi r^2 \epsilon_s(r)dr$ if we 
know the energy source distribution $\epsilon_s(r)$ in a cluster, which 
is approximately proportional to the sterile neutrino distribution. We 
demonstrate possible 
temperature profiles by using two models. In one we assume that the power 
sources are located at the centre of a cluster with $\epsilon_s(r)= 
\delta(r)L/4 \pi r^2$, where $L$ is the total 
luminosity of the cluster. In the other one, we assume that 
the power sources are distributed as $\epsilon_s=Ae^{-r^2/r_c^2}$, 
where $A$ can be fixed by the total luminosity $L$. Fig.~4 shows the 
temperature profiles of the two models. We can see that the 
temperature 
varies only within 6 percents of the central temperature 
$T_0$, which is due to the large value of $K_c$. 
Therefore, the energy is transferred from the sources (decayed $m_{s2}$ 
neutrinos) to the hot gas quite efficiently within the Hubble time.

\section{Discussion and Summary}
It has been a puzzle
why some clusters are still so hot even though their cooling times are 
shorter than the Hubble time. Even though cooling 
flow models \cite{xray} can help to explain this phenomenon, significant 
cooling flow 
in clusters has not been observed. This is known as the `mass sink 
problem' \cite{coolingflow}. 
Also, the cooling flow model predicts that the temperature of the 
hot gas should be quite inhomogeneous. However, 
the observed hot gas temperature profiles are quite homogeneous and gas 
with temperature less than 1 keV does not exist in the amount predicted. 
Donahue and Voit summarized the drawbacks of the cooling flow models and 
suggested that any successful model must explain the mass sink problem 
and the homogeneous temperature profile with positive core temperature 
gradients extending to $\sim 10^2$ kpc \cite{coolingflow}. 
Here, we present an alternative solution 
of the mass sink problem. We have discussed how the radiative decay of 
sterile neutrinos can heat up the hot gas in clusters. Our model predicts 
a redshift dependence of $L_x/M$, Eq.~(15), which is consistent with 
the observed data
(see Fig.~1), and we obtain the decay rate of the sterile 
neutrinos in clusters, which is about $\Gamma \sim (4.6 \pm 1.1) \times 
10^{-17}$ s$^{-1}$. Our picture does not require any cooling flow, and the 
radiative heating power by neutrino decays results in a quite 
homogeneous temperature profile. Therefore, our model can fit the observed 
information and solve the cooling flow problem for a range of $\Gamma$. 

If we can constrain the values of $m_1$ and $\alpha_1$ \cite{m1}, 
we can get a better picture and fix $\Gamma$. The radius of a $10^{13}$ 
solar mass $m_{s1}$ 
neutrino star is of order 10 kpc which is the same scale as a 
galaxy. Therefore we propose 
that the galactic dark matter may be composed of $m_{s1}$ and 
$m_{s2}$ sterile neutrinos. By examining the rotation curve of a galaxy, 
we can obtain a feasible range of $m_{s1}$. 
If $m_1$ is below 1 eV, the size of an $m_1$ neutrino star is greater 
than 1 Mpc, which is the cluster scale. Therefore, the active neutrinos 
may affect the properties of hot gas and clusters \cite{manho}. 
Different neutrinos may correspond to structures in different scales. 
It would be interesting to study whether our model can solve both the dark 
matter problem and the cooling flow problem.

\section{Appendix}
We calculate how the Compton scattering of photons and 
electrons can heat up the hot gas in clusters.
In the usual Compton scattering calculation, the electron is at rest. 
However, the 
temperature of the hot gas in most clusters are at 
$10^7-10^8$ K, and so we should 
generalize the situation by transforming to the electron's rest frame 
\cite{lorentz}.
Therefore, we have
\begin{equation}
\epsilon'= \gamma \epsilon(1- \beta \cos \theta),
\end{equation}
\begin{equation}
\epsilon_s= \gamma \epsilon'_s(1+ \beta \cos \theta'_s),
\end{equation}
where $\epsilon$ and $\epsilon'$ are the initial energy of the photon in 
the lab 
frame and electron's rest frame respectively,
and $\theta$ is the angle between the initial photon direction and 
electron velocity in the lab frame. The subscript `s' 
indicates the scattered energy or angle. The scattered power in the rest 
frame is given by
\begin{equation}
P'=c \sigma \int \epsilon'_s f(p'_s)d^3p'_s,
\end{equation}
where $\sigma$, $f$ and $p'$ are the Compton cross section, distribution 
function and momentum of photons respectively.
Since the distribution of the photons is 
not necessarily isotropic, we need to transform back to the lab frame 
and integrate over all $\theta'_s$. Making use of
$d^3p'=d^3p(1- \beta \cos \theta)$, and combining Eqs.~(25) 
and (26), we have
\begin{equation}
P'=c \sigma \int \gamma^2(1- \beta \cos \theta_s)^2 \epsilon_s 
f(p_s)d^3p_s.
\end{equation}
After summing up all the momenta of the photon, we have
\begin{equation}
P'=c \sigma \gamma^2 \left(1+ \frac{\beta^2}{3} \right) \int^{ \infty}_{0} 
\epsilon_s n( \epsilon_s)d 
\epsilon_s=c \sigma \gamma^2 \left(1+ \frac{\beta^2}{3} \right)U_{ph_s},
\end{equation}
where $U_{ph_s}$ is the scattered energy density of the photons in the lab 
frame. Since 
$P'=\frac{dE'}{dt'}$ and $<dE_s>=\gamma <dE'_s>$, we have $P'=P$, 
where $P$ is the scattered power 
in the lab frame. In the lab frame, the rate of energy removed from the 
photon field is
\begin{equation}
P= \frac{dE}{dt}= -c \sigma \int \epsilon n( \epsilon)d \epsilon=-c 
\sigma U_{ph},
\end{equation}
where $U_{ph}$ is the energy density of the photons in the lab frame. 
Therefore, 
the net energy radiated is the difference 
of energy scattered and energy removed in lab frame:
\begin{equation}
P_{net}=c \sigma \gamma^2 \left(1+ \frac{\beta^2}{3} \right)U_{ph_s}-c 
\sigma U_{ph}.
\end{equation}
If all the photons undergo Thomson scattering with electrons, then 
$\epsilon'=\epsilon'_s$ in Eq.~(30) and 
$U_{ph}$ will be the same as $U_{ph_s}$. However, in general, $\epsilon'$ 
may not be equal to $\epsilon'_s$. 
Suppose we have only a particular energy for the photon field, 
$U_{ph}=n_{ph} 
\epsilon$. From Eq.~(30), we have
\begin{equation}
P_{net}=c \sigma n_{ph} \left[ \gamma^2 \left(1+ \frac{\beta^2}{3} 
\right)(\epsilon - \Delta \epsilon)- \epsilon \right],
\end{equation}
where $\Delta \epsilon= \epsilon - \epsilon_s$. After simplification, 
Eq.~(31) will become
\begin{equation}
P_{net}=c \sigma n_{ph} \gamma^2 \left[ \frac{4}{3} \beta^2 \epsilon^2- 
\left(1+ \frac{ \beta^2}{3} \right) \Delta 
\epsilon \right].
\end{equation}
From Eqs.~(24) and (25), we have
\begin{equation}
< \Delta \epsilon'>= \epsilon'- \epsilon'_s= \gamma \Delta \epsilon.
\end{equation}
Therefore, the net power radiated in the lab frame is
\begin{equation}
P_{net}=c \sigma n_{ph} \gamma^2 \left[ \frac{4}{3} \beta^2 \epsilon- 
\left(1+ \frac{ \beta^2}{3} \right) 
\frac{< \Delta \epsilon'>}{\gamma} \right].
\end{equation}
If the net power radiated is a positive value, that means there is some 
energy removed from the electron gas 
and thus the energy of photon field is increased. This is known as Inverse 
Compton Scattering. However, it is possible 
that the value of the net power radiated is negative, which 
means that the electrons absorb energy from the 
photon field. If the photon and electron Compton scatter in 
the rest frame of the electron, then we have
\begin{equation}
\Delta \epsilon'= \frac{ \epsilon'^2}{m_ec^2}(1- \cos \phi)= \frac{ 
\gamma^2 \epsilon^2}{m_ec^2}(1- \beta \cos \theta)^2
(1- \cos \phi).
\end{equation}
We integrate over all the angles to get the averaged energy gain in 
one scattering:
\begin{equation}
<\Delta \epsilon'>= \frac{\gamma^2 \epsilon^2}{m_ec^2} \left(1+ \frac{ 
\beta^2}{2} \right).
\end{equation}
Combining Eqs.~(32), (33) and (36), we get
\begin{equation}
P_{net}=c \sigma n_{ph} \gamma^2 \left[ \frac{4}{3} \beta^2 
\epsilon- \frac{ \gamma \epsilon^2}{m_ec^2} \left(1+ \frac{ \beta^2}{3} 
\right) \left(1+ \frac{ \beta^2}{2} \right) \right].
\end{equation}
Therefore, the criterion for energy absorption from the photon field is
\begin{equation}
\epsilon> \frac{4m_ec^2 \beta^2}{3 \gamma \left(1+ \frac{ \beta^2}{3} 
\right) \left(1+ \frac{ \beta^2}{2} \right)}.
\end{equation}
Since $< \beta^2>= 
\frac{3kT}{m_ec^2}$, we can write 
$ \epsilon$ in terms of $kT$.  For non-relativistic electron, $\beta<<1$, 
$\epsilon > 4kT/ \gamma$. So, if the photon 
energy is greater than the kinetic energy of the electron, the electron 
will gain energy and vice versa.

\begin{table}[t]
\caption{X-ray Luminosities($L_x$), hot gas masses($M_g$), total 
masses($M_{\rm{total}}$) and redshifts($z$) of 12 clusters [1], [2].}
\label{table}
\end{table}
\begin{center}
\begin{tabular}{|c|c|c|c|c|}
\hline
Cluster &$L_x$(0.01-40 keV)(10$^{44}$ ergs$^{-1})$ &$M_g$ 
(10$^{14}$ $M_{\odot}$) & $M_{\rm{total}}$ (10$^{14}$ $M_{\odot}$) & $z$  
\\
\hline
A 119 &3.814 &0.42 &8.74 &0.0440  \\
\hline
A 133 &2.749 &0.17 &4.79 &0.0569  \\
\hline
A 399 &9.083 &1.2 &11.6 &0.0715  \\
\hline
A 401 &17.38 &0.64 &11.58 &0.0748  \\
\hline
A 754 &6.106 &0.63 &15.67 &0.0528  \\
\hline
A 539 &0.987 &0.08 &4.31 &0.0288  \\ 
\hline
A 1367 &1.092 &1.30 &5.77 &0.0216  \\
\hline
A 1775 &2.926 &1.20 &5.86 &0.0757  \\
\hline
A 2065 &6.261 &0.53 &14.44 &0.0721  \\
\hline
A 2255 &6.999 &3.45 &12.53 &0.0800  \\
\hline
A 2256 &11.588 &3.65 &13.81 &0.0601  \\
\hline
A 2634 &0.930 &2.20 &5.55 &0.0312  \\
\hline
\end{tabular}
\end{center}

\begin{figure}[h]
\centerline{\epsfig{file=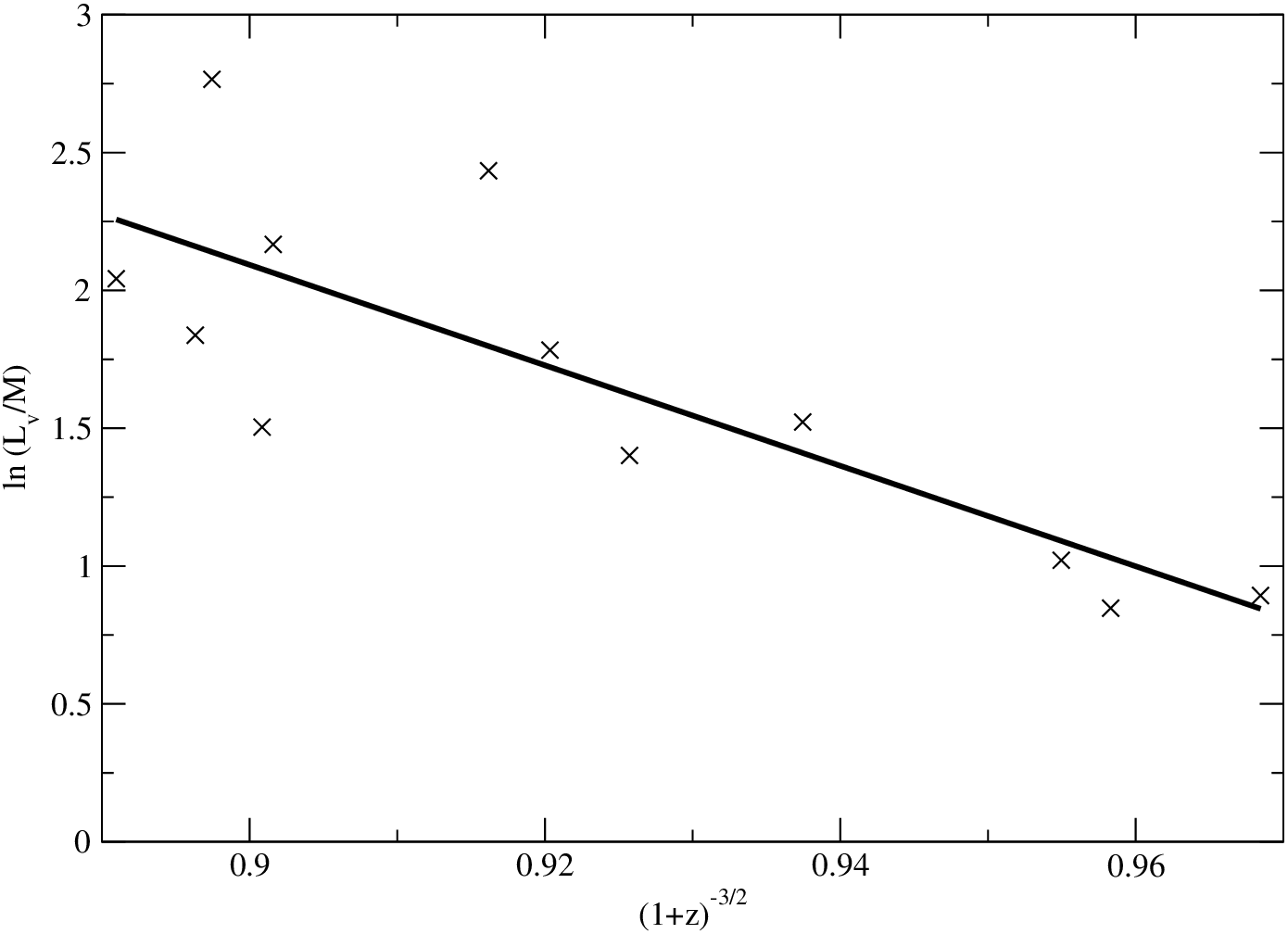, angle=0, width=12cm}}
{Fig.1. $\ln (L_x/M)$ vs. $(1+z)^{-3/2}$  
for 12 different clusters. The solid line is the best fit line with 
slope 
$-18.23 \pm 4.19$.}
\end{figure}
\vskip 10mm 

\begin{figure}
\centerline{\epsfig{file=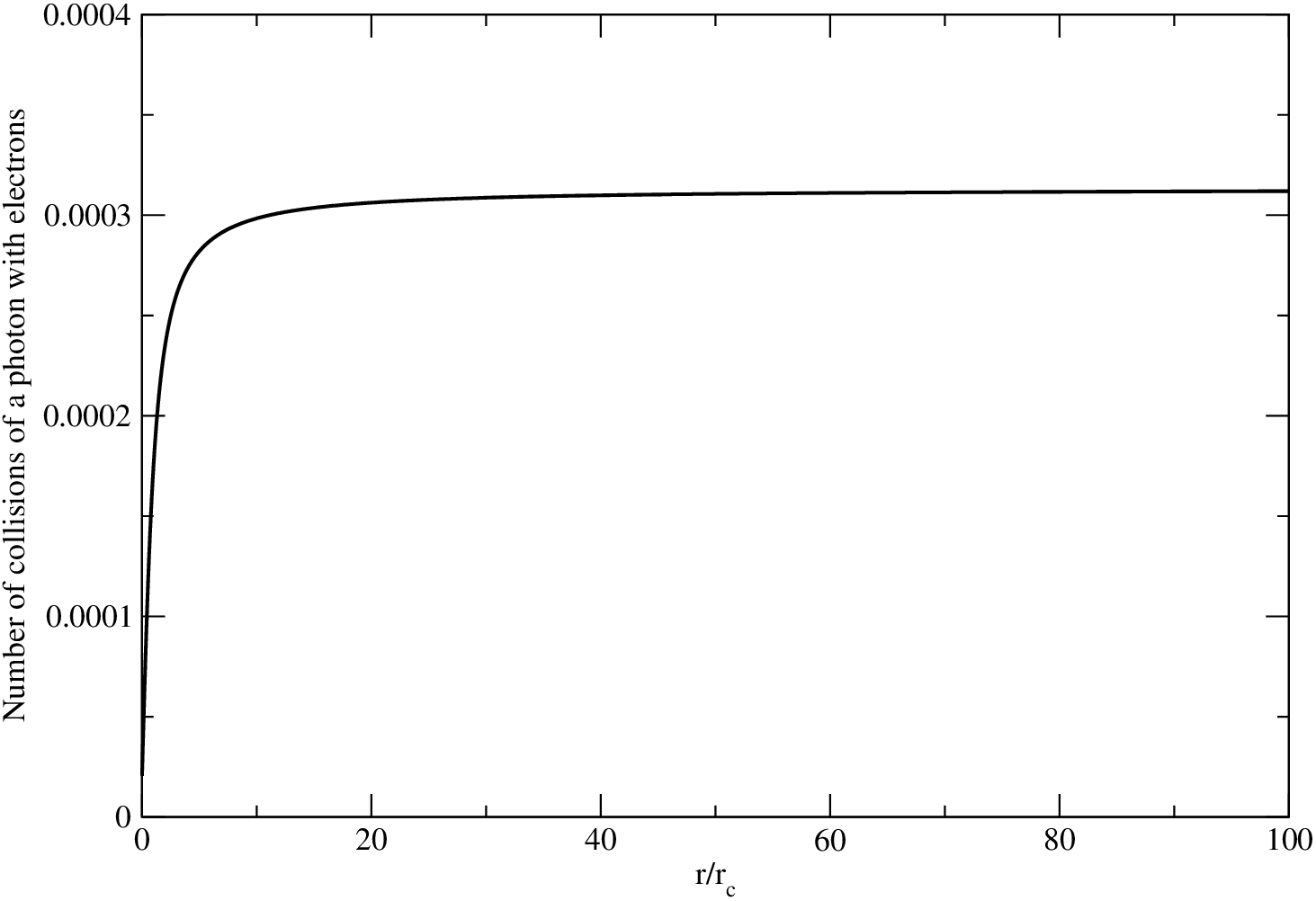, angle=0, width=12cm}}
{Fig.2. Number of collisions of a photon with electrons 
vs.~distance from the cluster centre. We assume $r_c=100$ kpc, 
$n_c=10^{-3}$ cm$^{-3}$ and $\beta_c=0.7$.}
\end{figure}
\vskip 10mm 

\begin{figure}
\centerline{\epsfig{file=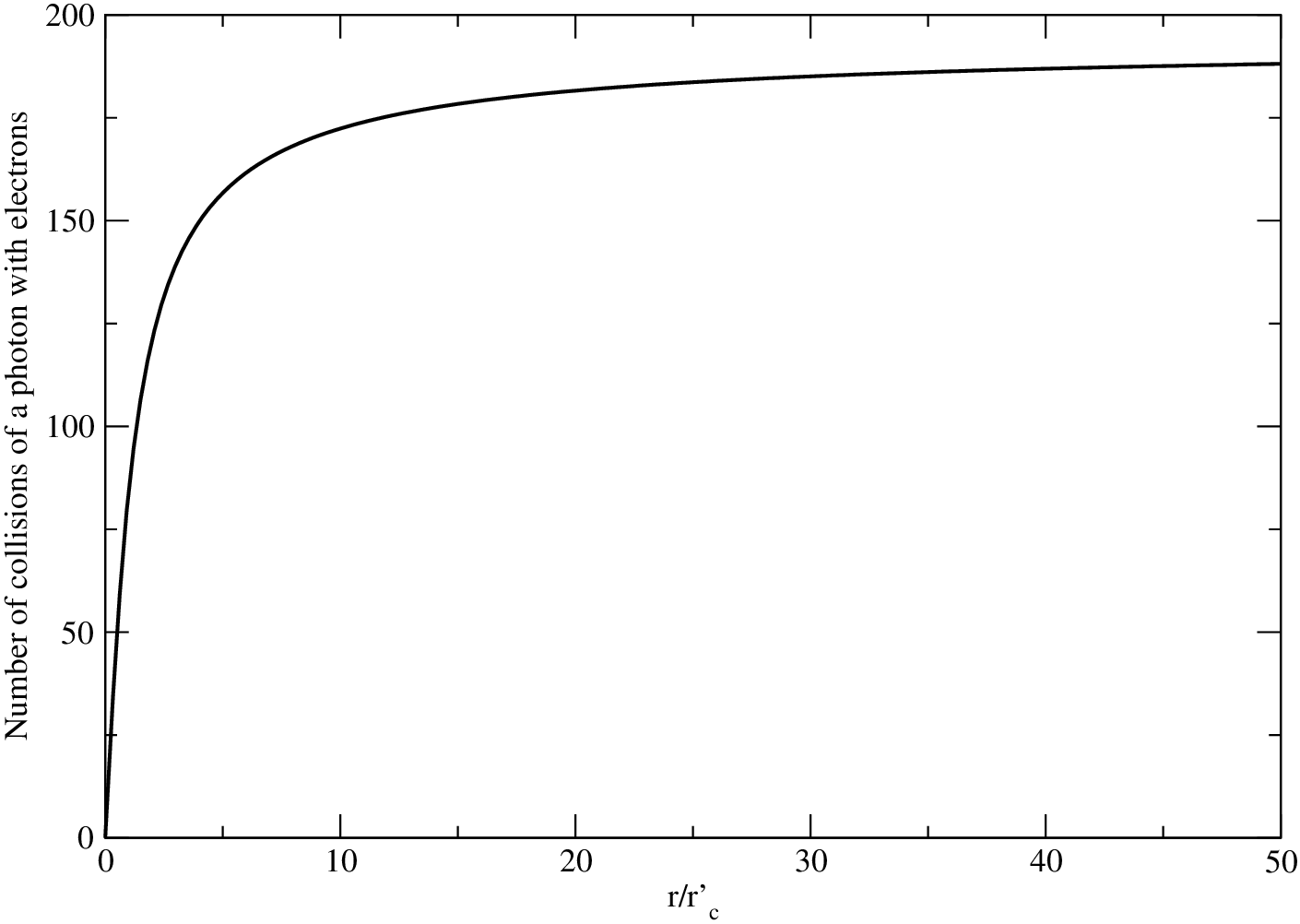, angle=0, width=11cm}}
{Fig.3. Number of collisions of a photon with electrons  
in Milky Way with $r_c'=0.34$ pc and $n_c'=1.6 \times 10^8$ cm$^{-3}$.}
\end{figure}
\vskip 5mm 

\begin{figure}
\centerline{\epsfig{file=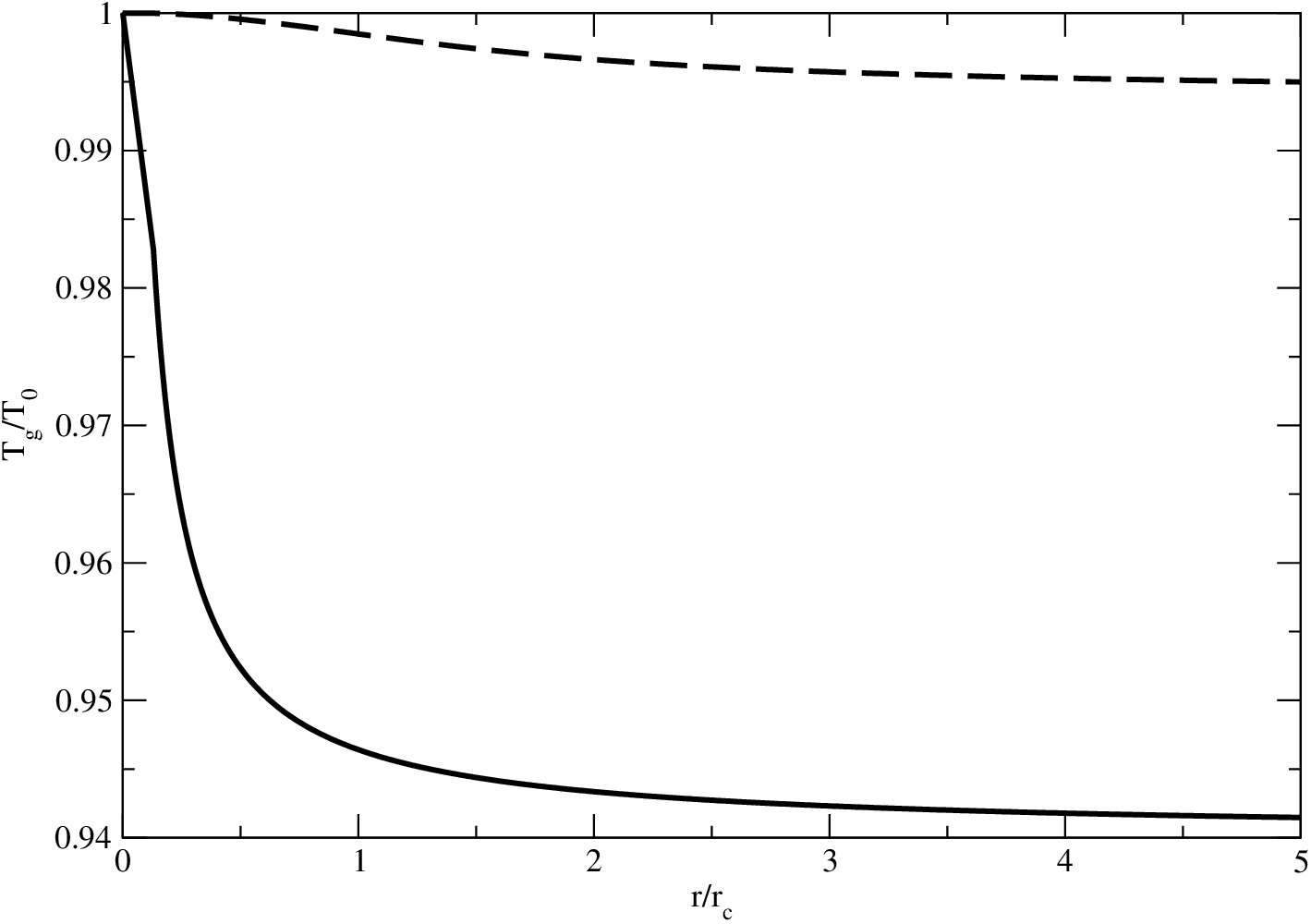, angle=0, width=11cm}}
{Fig.4. The temperature profiles of the two models discussed in the text. 
The solid line 
represents the model with the sources at the centre of the cluster. The 
dashed line represents the model 
with $\epsilon (r)=Ae^{-r^2/r_c^2}$. We assume $L=10^{44}$ ergs s$^{-1}$, 
$r_c=100$ kpc, and the central temperature $T_0=10^8$ K.}
\end{figure}

\end{document}